\documentclass[12pt]{iopart}

\newcommand{\mbf}[1]{\mbox{\boldmath$#1$}}

\newcommand{\dk}[0]{\int \overline{dk} }
\newcommand{\GS}[1]{#1\!\!\!\!\!\not~} 

\newcommand{\kizilersu}{K{\i}z{\i}lers{\" u} }

\usepackage{graphicx}

\begin{document}

\title[Checking the transverse Ward-Takahashi relation at one loop order]{Checking the transverse Ward-Takahashi relation at one loop order in 4-dimensions}

\author{M R Pennington and R Williams}
\address{Institute for Particle Physics Phenomenology,\\ Durham University, \\Durham DH1 3LE, U.K.}
\ead{\mailto{m.r.pennington@durham.ac.uk}, \mailto{richard.williams@durham.ac.uk}}

\begin{abstract}
Some time ago Takahashi derived so called {\it transverse} relations relating Green's functions of different orders to complement the well-known Ward-Green-Takahashi identities of gauge theories by considering wedge rather than inner products. These transverse relations have the potential to determine the full fermion-boson vertex in terms of the renormalization functions of the fermion propagator. He \& Yu have given an indicative proof at one-loop level in 4-dimensions. However, their construct involves the 4th rank Levi-Civita tensor defined only unambiguously in 4-dimensions exactly where the loop integrals diverge. Consequently, here we explicitly check the proposed transverse Ward-Takahashi relation holds at one loop order in $d$-dimensions, with $d=4+\varepsilon$.
\end{abstract}
\submitto{\jpg}
\maketitle

\section{Introduction}
   The Ward-Green-Takahashi identities~\cite{Ward:1950xp,Green:1953te,Takahashi:1957xn} play an important role in the study
of gauge theories and particularly in the implementation of consistent non-perturbative truncations of the corresponding Schwinger-Dyson equations~\cite{Ball:1980ay,Curtis:1990zs}.
The Ward-Green-Takahashi identities involve contractions of vertices
with external momenta and relate these to Green's functions with a lesser number of external legs. The best known of these relates the 3 point vector vertex
coupling a fermion-antifermion pair to the gauge boson, $\Gamma^{\mu}_V(p_1,p_2)$, to the
difference of fermion propagators, $S_F(p_1)$ and $S_F(p_2)$, so that with $q\,=\,p_1\,-\,p_2$
\begin{equation}\label{eqn:wgti}
  	q_{\mu}\,\Gamma^{\mu}_V (p_1,p_2)\;=\;S_F^{-1}(p_1)\,-\,S_F^{-1}(p_2)\; .
\end{equation}
Such projections constrain the so called longitudinal component of the vertices, while leaving their transverse parts unrestricted. 
\begin{figure}[b]
\vspace{3mm}
\begin{center}
\includegraphics[width=5.5cm]{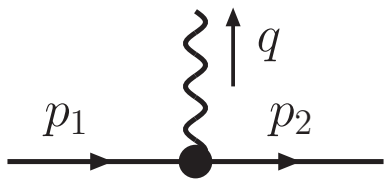}
\end{center}
   \caption{\small{Fermion-vector boson vertex and its momenta}}
\end{figure}

\noindent While the Ward-Green-Takahashi identity follows from the divergence of the vector vertex, Takahashi made plausible the existence of new relations that follow from the {\it curl} (or wedge product) of the vertex~\cite{Takahashi:1985yz}. These are referred to as transverse Ward-Takahashi relations and have the potential to restrict the transverse vertices from gauge symmetry alone. Kondo~\cite{Kondo:1996xn} rederived these relations for the 3-point functions in coordinate space using the path integral formulation. Subsequently, He, Khanna and Takahashi~\cite{He:2000we} using canonical field theory cast these relations in momentum space.
However, He~\cite{He:2002jg} then showed that an essential part of the Fourier transform was overlooked and that the correct relation in QED (in the simpler massless fermion) case is:
\begin{eqnarray}\label{eqn:twtr}
	iq^\mu\Gamma^\nu_V-iq^\nu\Gamma^\mu_V&=&S^{-1}_F(p_1)\sigma^{\mu\nu}+\sigma^{\mu\nu}S^{-1}_F(p_2)-\frac{1}{2}(p_1+p_2)_\lambda \left\{ \sigma^{\mu\nu}, \Gamma^\lambda_V \right\}\nonumber\\[2.7mm]
	&&+\int\frac{d^dk}{\left( 2\pi \right)^d}\;k_\lambda\,\left\{\sigma^{\mu\nu}, \tilde\Gamma_V^\lambda\right\}\;,
\end{eqnarray}
%\begin{eqnarray}\label{eqn:twtr}
%  	iq^\mu\Gamma^\nu_V(p_1,p_2)-iq^\nu\Gamma^\mu_V(p_1,p_2)&=&S^{-1}_F(p_1)\sigma^{\mu\nu}+\sigma^{\mu\nu}S^{-1}_F(p_2)+(p_1+p_2)_\lambda\varepsilon^{\lambda\mu\nu\rho}\Gamma_{A\rho}(p_1,p_2)\nonumber\\[2.7mm]
%  	&&-\int\frac{d^dk}{\left( 2\pi \right)^d}\;2k_\lambda\,\varepsilon^{\lambda\mu\nu\rho}\,\tilde\Gamma_{A\rho}(p_1,p_2;k)\;,
%\end{eqnarray}
\noindent where $\sigma^{\mu\nu}\,=\,i\,\left[ \gamma^\mu,\gamma^\nu \right]/2$, $\Gamma^\mu_V=\Gamma^\mu_V(p_1,p_2)$ and $\tilde\Gamma_V^\lambda=\tilde\Gamma_V^\lambda(p_1,p_2;k)$. In contrast to what is written in the paper by He~\cite{He:2002jg}, it is the anticommutator $\left\{ \sigma^{\mu\nu},\gamma^\lambda \right\}$ that is used in the derivation of Kondo~\cite{Kondo:1996xn}; the identification of $\left\{ \sigma^{\mu\nu},\gamma^\lambda \right\}$ with $-2\varepsilon^{\lambda\mu\nu\rho}\gamma_\rho\gamma_5$ is only valid in exactly $4$ space-time dimensions and should not be used when the requisite integrals contain divergences. It is this requirement that motivates the analysis presented here. 

\noindent The last two terms of (\ref{eqn:twtr}) are given by the momentum transform of 
\begin{equation}\label{eqn:wilsonlinecoord}
	\lim_{x'\rightarrow x}-\frac{i}{2}\left(\partial_\lambda^x-\partial_\lambda^{x'}\right)\left< 0\right| T\bar{\psi}(x')\left\{ \sigma^{\mu\nu},\gamma^\lambda \right\} U_P(x',x)\psi(x)\psi(x_1)\bar{\psi}(x_2)\left| 0 \right>
\end{equation}
with 
\begin{equation}
  	U_P(x',x)\,=\,\mathcal{P} \exp\left(-ig\int_x^{x'} dy^\rho A_\rho(y)\right)\; ,
\end{equation}
where $\psi$ and $A$ are the fermion and gauge fields respectively. $\mathcal{P}$ means the integral is path ordered, in fact a Wilson line integral. This implicitly defines the non-local vector vertex in our anticommutator $\left\{\sigma^{\mu\nu},\tilde\Gamma_V^\lambda\right\}$ of (\ref{eqn:twtr}), related to the axial one $\tilde\Gamma_{A\rho}(p_1,p_2;k)$ in He~\cite{He:2002jg}.
Being   non-local these involve integrals in momentum space
some of which cannot be represented diagrammatically in terms of Feynman graphs. An example is the last term of (\ref{eqn:twtr}), which for later we denote by $\mbf{W}^{\mu\nu}$. With a textbook factor of $Z_1$ renormalising the vector vertices and $Z_2^{-1}$ renormalising the fermion propagator, multiplicative renormalisability of the transverse Ward-Takahashi relation, (\ref{eqn:twtr}), is ensured by the same condition $Z_1 = Z_2$ as required by the renormalisation of the longitudinal Ward-Green-Takahashi identity, (\ref{eqn:wgti}).

\noindent To see how the Ward-Takahashi identity and the transverse relation  constrain the fermion-boson vector vertex, first
let us recall the very first Ward identity, which is the $q\,\to\,0$ limit of (\ref{eqn:wgti}), {\it viz.}
\begin{equation}\label{eqn:ward}
\Gamma^{\mu}_V(p,p)\;=\;\frac{\partial S_F^{-1}(p)}{\partial p_{\mu}}\quad .
\end{equation}
Let us separate 
the vertex into longitudinal and transverse components defined by
\begin{equation}
\Gamma^{\mu}_V\,=\, \Gamma_{L}^{\mu}\,+\,\Gamma_{T}^{\mu}\qquad ,\quad {\rm with}\quad q_{\mu}\,\Gamma_T^{\mu}\,=\,0 \quad .
\end{equation}
We can arrange for $\Gamma_L^{\mu}$ alone to satisfy the original Ward-Green-Takahashi identity, (\ref{eqn:wgti}), and for $\Gamma_T^{\mu}$ alone to contribute to the left hand side of the transverse Ward-Takahashi relation, (\ref{eqn:twtr}), by writing
\begin{equation}\label{eqn:singularsplitting}
\Gamma^{\mu}_L\,=\,\frac{q^{\mu}}{q^2}\,\left(S_F^{-1}(p_1)\,-\,S_F^{-1}(p_2)\right)\quad , \, {\rm and}\quad \Gamma_T^{\mu}\,=\,\left( g^{\mu\nu} - \frac{q^{\mu}q^{\nu}}{q^2}\right)\,\Gamma_V^{\mu} \; .
\end{equation}
This separation seemingly makes the longitudinal and transverse components unrelated.
However, each component of (\ref{eqn:singularsplitting}) has a kinematic singularity at $q^2 = 0$. Consequently, the Ward identity in (\ref{eqn:ward}) requires an inter-relation between $\Gamma_{L}^{\mu}$  and
$\Gamma_{T}^{\mu}$. An alternative separation is to abandon (\ref{eqn:singularsplitting}) and ensure that each component is free of kinematic singularities. Then one can require that the longitudinal part alone not only satisfies (\ref{eqn:wgti}), but its $q \to 0$ limit too, {\it viz.} (\ref{eqn:ward}). This is the Ball-Chiu construction~\cite{Ball:1980ay}. With this definition of the longitudinal part, we have, writing the massless inverse fermion propagator as
$S_F(p)\,=\,\alpha(p^2) \GS{p}$:
\begin{equation}\fl
\label{eqn:ball-chiu}
\Gamma^{\mu}_L(p_1,p_2)\;=\;\frac{1}{2}\,\left (\alpha(p_1^2)+\alpha(p_2^2)\right)\,\gamma^{\mu}\;+\;\frac{1}{2}\,\frac{\alpha(p_1^2)-\alpha(p_2^2)} {p_1^2-p_2^2}\,
(\GS{p}_1+\GS{p}_2) (p_1+p_2)^{\mu}\quad .
\end{equation}
The transverse component is then only constrained to satisfy the condition: $\Gamma^{\mu}_T(p,p)\,=\,0$, which the separation of (\ref{eqn:singularsplitting}) does not.
 However, the {\it transverse} Ward-Takahashi relation now involves both $\Gamma^{\mu}_T$ and $\Gamma^{\mu}_L$ as well.
We can illustrate the power of the transverse relation by considering
the $q \to 0$ limit of this equation. We can then deduce to all orders in perturbation theory, and genuinely non-perturbatively, the constraint
$\mbf{W}^{\mu\nu}(p+q,p)$ places on the transverse vertex. From (\ref{eqn:twtr}) we have:
\begin{eqnarray}\fl
\mbf{W}^{\mu\nu}(p_1,p_2)&\equiv&
\int\frac{d^dk}{\left( 2\pi \right)^d}\;k_\lambda\,\left\{ \sigma^{\mu\nu},\tilde\Gamma_V^\lambda(p_1,p_2;k)\right\}\nonumber\\[3.5mm]
&=&-\Biggl\{S^{-1}_F(p_1)\sigma^{\mu\nu}+\sigma^{\mu\nu}S^{-1}_F(p_2)-\frac{1}{2}(p_1+p_2)_\lambda\left\{\sigma^{\mu\nu},\Gamma_V^\lambda(p_1,p_2)\right\}\nonumber\\[2.mm]
&&{\hspace{23mm}}-	iq^\mu\Gamma^\nu_V(p_1,p_2)+iq^\nu\Gamma^\mu_V(p_1,p_2) \Biggr\}\,.
\end{eqnarray}
When $q \to 0$ then $\mbf{W}^{\mu\nu}(p_1=p_2=p)\,=\,0$.
The general transverse vertex in the massless fermion case involves 4 vectors orthogonal to $q^{\mu}$, the basis vectors $T_i^{\mu}$, which are listed in Ref.~\cite{Kizilersu:1995iz}, so that
\begin{equation}
\Gamma^{\mu}_T(p+q,p)\;=\;\sum_{i=2,3,6,8}\, \tau_i((p+q)^2,p^2,q^2)\, T_i^{\mu}(p,q)\quad ,
\end{equation}
where the coefficients $\tau_i$ are themselves free of kinematic singularities and are functions of the three relevant invariants, $p_1^2$, $p_2^2$ and $q^2$.
$T_2^{\mu}$ and $T_3^{\mu}$ are quadratic in $q$, while $T_6^{\mu}$ and $T_8^{\mu}$ are both linear. In particular 
\begin{equation}
	T_8^{\mu}\; =\; i \gamma^{\mu} p_1^{\nu} p_2^{\lambda} \sigma_{\nu\lambda}\,+p_1^{\mu} \GS{p}_2\,-\,p_2^{\mu} \GS{p}_1\; .
\end{equation}
The coefficients $\tau_i$ are all symmetric under $p_1=p+q \leftrightarrow p_2=p$, except for $\tau_6$, which is antisymmetric and so vanishes when $q \to 0$.
 We then have non-perturbatively to 
first order in the boson momentum~$q$:
\begin{eqnarray}\fl 
\mbf{W}^{\mu\nu}(p_1 = p+q, p_2=p)&=&-i\left[ \left(p^{\mu} q^{\nu}\,-\,p^{\nu} q^{\mu}\right)\,\GS{p}\,+\,q \cdot p\,\left(p^{\mu} \gamma^{\nu}\,-\,p^{\nu} \gamma^{\mu}\right) \right]\,\alpha'(p^2)\nonumber\\[3.mm]
&+&2i\left[ \left(p^{\mu} q^{\nu}\,-\,p^{\nu} q^{\mu}\right)\,\GS{p}\,-\,q \cdot p\,\left(p^{\mu} \gamma^{\nu}\,-\,p^{\nu} \gamma^{\mu}\right) \right.\nonumber\\[3.mm]
&&\hspace{6mm}+\left. p^2\left( q^\mu\gamma^\nu-q^\nu\gamma^\mu \right)\right]\tau_8\left( p,p \right) ,
%
%+\;\frac{1}{2}\,\tau_8(p,p)\,\left[ \GS{q} \gamma^{\mu} \GS{p}\,-\,\GS{p} \gamma^{\mu} \GS{q} \right] ,
\end{eqnarray}
where we expect $\alpha^{\prime}(p^2)$ to result from terms like
$\left(\alpha(p_1^2) - \alpha(p_2^2)\right)/\left(p_1^2 - p_2^2\right)$ as in (\ref{eqn:ball-chiu}). Such constraints restrict the form of the transverse coefficient $\tau_8$, for example, and have the potential to play a critical role in constructing
consistent non-perturbative Feynman rules. Consequently, it is critical to check how to evaluate the transverse Ward-Takahashi relation, which is what we do here.

\noindent It is trivial to check that the {transverse Ward-Takahashi relation} holds at tree level.
This follows directly from the identity
\begin{equation}\label{eqn:tlidentity}
	i q^\mu\gamma^\nu-i q^\nu\gamma^\mu=\GS{p}_1\sigma^{\mu\nu}+\sigma^{\mu\nu}\GS{p}_2-\frac{1}{2}(p_1+p_2)_\lambda\left\{ \sigma^{\mu\nu},\gamma^\lambda \right\}\; .
\end{equation}
when the non-local term does not appear. Note that at tree level we are in $4$-dimensions so we would be permitted to use the identity $\left\{ \sigma^{\mu\nu},\gamma^\lambda \right\}=-2\varepsilon^{\lambda\mu\nu\rho}\gamma_\rho\gamma_5$. He and Yu \cite{He:2003fa} have sketched a proof of the relation at one loop order. The purpose of this paper is to check this in detail in $d$-dimensions, where $d=4+\epsilon$. 
%While the original Ward-Green-Takahashi identity, (\ref{eqn:wgti}), involves the vector vertex of the gauge theory,  the {transverse Ward-Takahashi relation} uses by construction the associated axial vector vertex, $\Gamma^{\mu}_A$. 
For notational brevity we introduce the following form of the Chisholm identity:
%%% put down ALL possible forms of writing this piece, notably the one's that exhibit symmetry as well
%\begin{eqnarray}\label{eqn:chisholm}
%r^{\lambda\mu\nu}&\equiv& 2\left( \gamma^\lambda\gamma^\mu\gamma^\nu - g^{\lambda\mu}\gamma^\nu +
%g^{\lambda\nu}\gamma^\mu - g^{\mu\nu}\gamma^\lambda \right)\nonumber\\[2.7mm]
%&\equiv&- i\left\{ \sigma^{\mu\nu},\gamma^\lambda\right\} \;=\; \left( \gamma^\lambda\gamma^\mu\gamma^\nu-\gamma^\nu\gamma^\mu\gamma^\lambda \right)\;.
%\end{eqnarray}
%
\begin{equation}\label{eqn:chisholm}
	- i\left\{ \sigma^{\mu\nu},\gamma^\lambda\right\} \;=\; \mbf{r^{\lambda\mu\nu}}\;=\;  \left( \gamma^\lambda\gamma^\mu\gamma^\nu-\gamma^\nu\gamma^\mu\gamma^\lambda \right)\,.
\end{equation}

\noindent There has been discussion in the literature~\cite{He:2001cu,Kondo:1996xn} about the role of the Adler-Bell-Jackiw anomaly~\cite{Adler:1969gk,Bell:1969ts,Adler:1969er} in these relations. It has been shown~\cite{Sun:2003jh,Sun:2003ia} that this plays no role in the axial-vector equivalent of (\ref{eqn:twtr}). 
We will consider the transverse Ward-Takahashi relation in the massless fermion case for simplicity. We show that this only holds if appropriate care is taken of the integrals divergent in 4-dimensions. This allows us to complete the proof to one loop order outlined by He and Yu~\cite{He:2003fa}.

\section{Perturbative Derivation of the Transverse\\  Ward-Takahashi Relation}

\noindent Following the formal derivation of the transverse Ward-Takahashi relation in Ref.~\cite{He:2000we}, He and Yu~\cite{He:2003fa} give a proof of how the relation should hold at one loop order by considering the relevant integrands, but without evaluating any of the integrals.
Here we will investigate the relation in greater detail. First in this section we will reconsider the proof by deducing the integrands in $d$-dimensions, where $d=4+\epsilon$. To manipulate the divergent integrals integrals in $d$-dimensions we must clearly use the original object $\left\{ \sigma^{\mu\nu},\gamma^\lambda \right\}$, since once again its identification with $-2\varepsilon^{\lambda\mu\nu\rho}\gamma_\rho\gamma_5$ is valid only in $4$-dimensions. We will then confirm this result in the next section by explicit evaluation of each of the contributing integrals.

%\newpage

\begin{figure}[b]
\begin{center}
\includegraphics[width=14.cm]{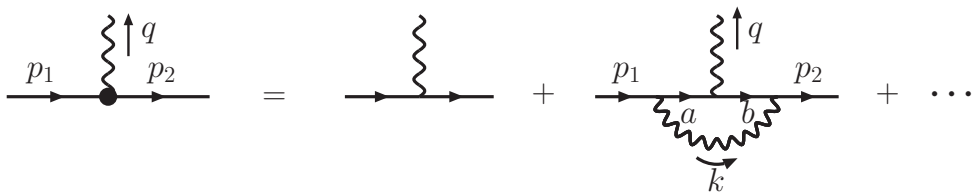}
\end{center}
\vspace{-3mm}
   \caption{\small{One loop corrections to the vector vertex of figure~1 with the momenta labelled as in our calculation}}
\end{figure}

\noindent To derive the transverse Ward-Takahashi relation at one-loop order, we begin with the tree-level relation of (\ref{eqn:tlidentity}):
\begin{equation}\label{eqn:rwtlidentity}
  	i q^\mu\gamma^\nu-i q^\nu\gamma^\mu\,=\,\GS{p}_1\sigma^{\mu\nu}+\sigma^{\mu\nu}\GS{p}_2-\frac{1}{2}\left[ \left( \GS{p}_1+\GS{p}_2 \right)\sigma^{\mu\nu}+\sigma^{\mu\nu}\left( \GS{p}_1+\GS{p}_2 \right) \right]\; .
\end{equation}
It is obvious that this relation is trivially satisfied at tree-level, hence we need only concentrate on the one-loop corrections to the relation. To simplify the answer we introduce the following shorthand for the integration:
\vspace{2.5mm}
\begin{equation}
  	\dk\;\equiv\; g^2\int\frac{d^d k}{\left( 2\pi \right)^d}\;\frac{-i}{k^2\,a^2\,b^2}\; \qquad ,
\end{equation}
\noindent where $a=p_1\,-\,k$ and $b=p_2\,-\,k$ as shown in figure~2. 
\noindent We use the standard Feynman rules for QED with, for example,
\begin{equation}
  	\Delta_{\alpha\beta}(k)=\frac{-i}{k^2}\left( g_{\alpha\beta}+\left( \xi-1 \right)\frac{k_\alpha k_\beta}{k^2} \right)
\end{equation}
for the bare gauge boson propagator, where $\xi$ is the usual covariant gauge parameter.
If we write the full vertex to one-loop order as
\begin{equation}\label{eqn:oneloopvertex}
  	\Gamma^\mu_V(p_1,p_2)\,=\,\gamma^\mu+\Lambda^\mu_{(2)}(p_1,p_2)\qquad ,
\end{equation}
we can read off from figure~2 that
\begin{equation}
  	\Lambda^\mu_{(2)}(p_1,p_2)\,=\,\dk\, \gamma^\alpha\GS{a}\gamma^\mu\GS{b}\gamma_\alpha\; 
\end{equation}
in the Feynman gauge, when $\xi =1$.
We see that the left-hand side of (\ref{eqn:twtr}) can be
   obtained by sandwiching (\ref{eqn:rwtlidentity}) between $\gamma^\alpha \GS{a}$ and $\GS{b}\gamma_\alpha$, then integrating over the measure $\dk$, so that:
\vspace{2.5mm} 
\begin{eqnarray}\label{eqn:interimcalc1}\fl
  	i q^\mu\Lambda^\nu_{(2)}\left( p_1,p_2 \right)-iq^\nu\Lambda^\mu_{(2)}\left( p_1,p_2 \right)&=&\dk \gamma^\alpha\GS{a}\left( \GS{p}_1\sigma^{\mu\nu}+\sigma^{\mu\nu}\GS{p}_2\right)\GS{b}\gamma_\alpha\nonumber\\[2.7mm]
  	&-&\frac{1}{2}\dk\gamma^\alpha\GS{a}\left[ \left( \GS{p}_1+\GS{p}_2 \right)\sigma^{\mu\nu}+\sigma^{\mu\nu}\left( \GS{p}_1+\GS{p}_2 \right) \right]\GS{b}\gamma_\alpha\; .
\end{eqnarray}

\noindent The second term on the right-hand side is the $-\frac{1}{2}(p_1+p_2)_\lambda\left\{\sigma^{\mu\nu}, \Lambda^\lambda_{V(2)}\right\}$ piece of the original relation and is one of the terms we wish to identify.

\begin{figure}[t]
%\vspace{-4mm}
\begin{center}
\includegraphics[width=14.cm]{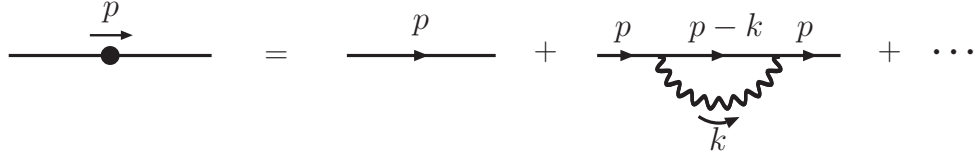}
\end{center}
\vspace{-4mm}
   \caption{\small{One loop correction to the fermion propagator}}
%\vspace{3mm}
\end{figure}

\noindent At the one-loop level we can write the fermion self energy parts of figure~3 as
\begin{eqnarray}\label{eqn:oneloopfermion}
  	S_F(p_1)&=&\GS{p}_1-\Sigma_{(2)}(p_1)\;,\qquad \textrm{where }\;\Sigma_{(2)}(p_1)=\dk\, \gamma^\alpha\left(\GS{a}b^2\right)\gamma_\alpha\nonumber\\[2.7mm]
  	S_F(p_2)&=&\GS{p}_2-\Sigma_{(2)}(p_2)\;,\qquad \textrm{where }\;\Sigma_{(2)}(p_2)=\dk\, \gamma^\alpha\left(a^2\GS{b}\right)\gamma_\alpha\;.
\end{eqnarray}

\noindent In order to extract such terms from (\ref{eqn:interimcalc1}), the unidentified term on the right hand side of this equation is re-expressed using the replacement:
\vspace{2.5mm}
\begin{eqnarray}\label{eqn:replacement}
\dk\, \gamma^\alpha\GS{a}\left(	\GS{p}_1\sigma^{\mu\nu}+\sigma^{\mu\nu}\GS{p}_2 \right)\GS{b}\gamma_\alpha&=&\dk\, \gamma^\alpha\left( 
-\sigma^{\mu\nu}a^2\GS{b}-\GS{a}b^2\sigma^{\mu\nu}\right)\gamma_\alpha\nonumber\\[2.7mm]
&&+\dk\gamma^\alpha\GS{a}\left(\GS{k}\sigma^{\mu\nu}+\sigma^{\mu\nu}\GS{k}\right)\GS{b}\gamma_\alpha\nonumber\\[2.7mm]
&&+ \dk\gamma^\alpha\GS{a}\left(2\GS{a}\sigma^{\mu\nu}+2\sigma^{\mu\nu}\GS{b}\right)\GS{b}\gamma_\alpha\;.
\end{eqnarray}

\noindent To the first line on the right-hand side of (\ref{eqn:replacement}) we commute the $\sigma$-matrices to the outside of the $\gamma^\alpha$ by using $\sigma^{\mu\nu}\gamma^\alpha\rightarrow \gamma^\alpha\sigma^{\mu\nu}-2ig^{\alpha\mu}\gamma^\nu+2ig^{\alpha\nu}\gamma^\mu$. 
\noindent This gives:
%\vspace{1.mm}
\begin{eqnarray}\label{eqn:beforegamma}\fl
  	i q^\mu\Lambda^\nu_{(2)}\left( p_1,p_2 \right)-iq^\nu\Lambda^\mu_{(2)}\left( p_1,p_2 \right)&=&-\,\Sigma_{(2)}(p_1)\,\sigma^{\mu\nu}-\sigma^{\mu\nu}\,\Sigma_{(2)}(p_2)\nonumber\\[3.5mm]
%	&-&\frac{1}{2}\dk\gamma^\alpha\GS{a}\left( \left( \GS{p}_1+\GS{p}_2 \right)\sigma^{\mu\nu}+\sigma^{\mu\nu}\left( \GS{p}_1+\GS{p}_2 \right) \right)\GS{b}\gamma_\alpha\nonumber\\
&&-\,\frac{1}{2}(p_1+p_2)_\lambda\left\{\sigma^{\mu\nu}, \Lambda^\lambda_{V(2)}\right\}\nonumber\\[3.5mm]
  	&&+\,\dk\gamma^\alpha\GS{a}\left( \GS{k}\sigma^{\mu\nu}+\sigma^{\mu\nu}\GS{k}\right)\GS{b}\gamma_\alpha\nonumber\\[3.5mm]
  	&&+2\dk \left\{ \gamma^\alpha\GS{a}\left( \GS{a}\sigma^{\mu\nu}+\sigma^{\mu\nu}\GS{b} \right)\GS{b}\gamma_\alpha\right.\nonumber\\[3.5mm]
  	&&\quad\left.-\left( \left( a^2\GS{b}+\GS{a}b^2 \right)\sigma^{\mu\nu}+\sigma^{\mu\nu}\left( a^2\GS{b}+\GS{a}b^2 \right) \right)\right\}	\; .
\end{eqnarray}

\noindent To be able to express this in terms of the Wilson line integral in the transverse 
Ward-Takahashi relation, we need to introduce a factor of $\gamma^\alpha$ and $\gamma_\alpha$ either side of the last term in (\ref{eqn:beforegamma}). On rearranging this becomes:
%%\newpage
\begin{eqnarray}\fl
  	i q^\mu\Lambda^\nu_{(2)}\left( p_1,p_2 \right)-iq^\nu\Lambda^\mu_{(2)}\left( p_1,p_2 \right)&=&-\Sigma_{(2)}(p_1)\,\sigma^{\mu\nu}-\sigma^{\mu\nu}\,\Sigma_{(2)}(p_2)\nonumber\\[3.mm]
&&-\,\frac{1}{2}(p_1+p_2)_\lambda\left\{\sigma^{\mu\nu}, \Lambda^\lambda_{V(2)}\right\}\nonumber\\[3.mm]
  	&&+\dk\,\gamma^\alpha\GS{a}\left( \GS{k}\sigma^{\mu\nu}+\sigma^{\mu\nu}\GS{k}\right)\GS{b}\gamma_\alpha\nonumber\\[3.mm]
  	&&-\dk\, \left( \gamma^\alpha\sigma^{\mu\nu}+\sigma^{\mu\nu}\gamma^\alpha \right)\left( a^2\GS{b}-\GS{a}b^2 \right)\gamma_\alpha\nonumber\\[3.mm]
  	&&-2(d-4)\dk\,\left( \GS{a}\sigma^{\mu\nu}+\sigma^{\mu\nu}\GS{a} \right)b^2\;.
\end{eqnarray}

\noindent Putting back in the tree-level result, together with the identification of last $3$ lines with the Wilson-line integration over non-local contributions  allows us to write the transverse Ward-Takahashi relation, derived for the massless case of QED in $d$-dimensions in the Feynman gauge, as
\vspace{3mm}
\begin{eqnarray}\label{eqn:twtrcorrect}\fl
  	iq^\mu\Gamma^\nu_V(p_1,p_2)-iq^\nu\Gamma^\mu_V(p_1,p_2)&=&S^{-1}_F(p_1)\,\sigma^{\mu\nu}+\sigma^{\mu\nu}\,S^{-1}_F(p_2)-\,\frac{1}{2}(p_1+p_2)_\lambda\left\{\sigma^{\mu\nu}, \Gamma^\lambda_V\right\}\nonumber\\[2.7mm]
	&&+\int\frac{d^dk}{\left( 2\pi \right)^d}\;k_\lambda\left\{\sigma^{\mu\nu},\tilde\Gamma_V^\lambda\right\}\;,  
\end{eqnarray}
which is indeed what is found by He, though written with explicitly $d$-dimensional objects. It is straightforward to check that this relation would hold in any covariant gauge.
\newpage
\section{The One Loop Integrals}

\noindent Since the integrals are not finite and so the manipulations more subtle, we 
now proceed to explicit computation of the terms in the transverse Ward-Takahashi relation to one loop order. The calculation will be divided into the four parts, of which it was originally composed, plus the fifth correction piece. 
The following sections deal with the terms in (\ref{eqn:twtrcorrect}) in turn
\begin{itemize}
\item {\bf section 3.1} ${\hspace{2cm}}i\,q^\mu\, \Gamma_V^\nu-i\,q^\nu\,\Gamma_V^\mu$
\vspace{1mm}
\item {\bf section 3.2} ${\hspace{2cm}}S^{-1}_F(p_1)\,\sigma^{\mu\nu}+\sigma^{\mu\nu}\,S^{-1}_F(p_2)$
\vspace{1mm}
\item {\bf section 3.3} ${\hspace{2cm}}\frac{1}{2}(p_1+p_2)_\lambda\;\left\{\sigma^{\mu\nu},\Gamma_V^\lambda\right\}$
\vspace{1mm}
\item {\bf section 3.4} ${\hspace{2cm}}\int d^4k\; k_\lambda\;\left\{\sigma^{\mu\nu},\tilde\Gamma_V^\lambda\right\}$
\end{itemize}
\vspace{1mm}

\noindent
The ${\cal O}(\alpha)$ correction to each of these we denote by $\left({i\alpha}\,{\bf P_i}^{\mu\nu}/4\pi\right)$ with $i=1,..,4$. Then (\ref{eqn:twtrcorrect}) if true would become simply
\begin{equation}\label{eqn:twtrassum}
{\bf P_1}^{\mu\nu}\;=\;\left[ {\bf P_2} + {\bf P_3} + {\bf P_4} \right]^{\mu\nu}\quad .
\end{equation}
This is what we set out to prove.
The calculation is performed in full for an arbitrary covariant gauge, specified as usual by $\xi$, and in dimension $d$. We will be particularly concerned with the results with $d\,=\,4 + \varepsilon$ when $\varepsilon\,\to\,0$.

\subsection{Part 1}
The first part of the calculation is the $\left(i\,q^\mu \Gamma^\nu_V-i\,q^\nu\Gamma^\mu_V\right)$ piece. This involves the full vertex $\Gamma^\mu_V(p_1,p_2)$, as calculated for arbitary covariant gauge by \kizilersu \emph{et al} \cite{Kizilersu:1995iz}. Once we know all of the relevant standard integrals, which are collected in the Appendix, the calculation is straightforward. We begin by writing the vertex to one-loop order as in (\ref{eqn:oneloopvertex})
\begin{equation}
  	\Gamma^\mu_V(p_1,p_2)=\gamma^\mu+\Lambda^\mu_{(2)}(p_1,p_2)\;.
\end{equation}

\noindent Defining the momentum flow from figure~2, we then have, using the standard Feynman rules, the vertex correction given by:
\begin{eqnarray}\fl
  	\Lambda^\mu(p_1,p_2)&=&-\frac{i\alpha}{4\pi^3}\int d^4\omega \frac{\gamma^\alpha\left( \GS{p}_1-\GS{\omega} \right)\gamma^\mu\left( \GS{p}_2-\GS{\omega} \right)\gamma^\beta}{\omega^2\left( p_1-\omega \right)^2\left( p_2-\omega \right)^2}\left( g_{\alpha\beta}+\left( \xi-1 \right)\frac{\omega_\alpha \omega_\beta}{\omega^2} \right)\; ,
\end{eqnarray}
\noindent
where we have suppressed the $+i\epsilon$ prescribed to define the propagators.
\noindent If we decompose this into its constituent tensor integrals,
\begin{eqnarray}\fl
\Lambda^\mu(p_1,p_2)=-\frac{i\alpha}{4\pi^3}\left\{ \gamma^\alpha\left( \GS{p}_1\gamma^\mu\GS{p}_2 \right)\gamma_\alpha\, \mbf{J^{(0)}}(p_1,p_2)\,-\,\gamma^\alpha\left( \GS{p}_1\gamma^\mu\gamma^\nu+\gamma^\nu\gamma^\mu\GS{p}_2 \right)\gamma_\alpha \,\mbf{J^{(1)}_\nu}(p_1,p_2)\right.\nonumber\\[4.0mm]
+\left( \gamma^\alpha\gamma^\nu\gamma^\mu\gamma^\lambda\gamma_\alpha \right)\mbf{J^{(2)}_{\nu\lambda}}(p_1,p_2)\,+\,\left( \xi-1 \right)\left[ \left( -\gamma^\nu\GS{p}_1\gamma^\mu-\gamma^\mu\GS{p}_2\gamma^\nu \right)\,\mbf{J^{(1)}_\nu}(p_1,p_2)\right. \nonumber\\[4.0mm]
\left. \left. +\gamma^\mu \mbf{K^{(0)}}(p_1,p_2)+\left( \gamma^\nu\GS{p}_1\gamma^\mu\GS{p}_2\gamma^\lambda \right)\mbf{I^{(2)}_{\nu\lambda}}(p_1,p_2) \right]\right\}\;.
\end{eqnarray}

\noindent Reducing the tensor integrals to scalar integrals, and collecting terms together with common Dirac structures gives
\begin{eqnarray}\fl
  	\Gamma^\mu_V\left( p_1,p_2 \right)&=& \mbf{\gamma^\mu}+\frac{\alpha}{4\pi}\left\{ \mbf{\GS{p}_2p^\mu_2}\left[ 2J_A-2J_C+\left( \xi-1 \right)\,I_D\, p_1^2\right] \right.\nonumber\\[2.0mm]
  	&+&\mbf{\GS{p}_1p_1^\mu}\left[ 2J_B-2J_E+\left( \xi-1 \right)\,I_D\, p_2^2\right]+\mbf{\GS{p}_1p_2^\mu}\left[ -2J_0+2J_A+2J_B-2J_D \right.\nonumber\\[2.0mm]
  	&&{\hspace{0.5cm}}+\left.\left( \xi-1 \right)\left( -\frac{1}{2} J_0 -\frac{1}{2} I_C\, p_2^2\,-\,I_D\, p_1\cdot p_2 +\frac{1}{2} I_E\, p_1^2 + J_A \right)\right]\nonumber\\[2.0mm]
  	&+&\mbf{\GS{p}_2p_1^\mu}\left[ -2J_D+\left( \xi-1 \right)\left( I_C\,p_2^2 - J_A \right) \right]+\mbf{\gamma^\mu\GS{p}_2\GS{p}_1}\left[ J_0-J_A-J_B \right.\nonumber\\[2.0mm]
  	&&{\hspace{0.5cm}}+\left.\left( \xi-1 \right)\left( \frac{1}{4} J_0 +\frac{1}{4} I_C\, p_2^2\,+\frac{1}{2} I_D\, p_1\cdot p_2 +\frac{1}{4} I_E\, p_1^2 - \frac{1}{2} J_A-\frac{1}{2} J_B \right)\right]\nonumber\\[2.0mm]
  	&+&\mbf{\gamma^\mu}\left[ - J_A\, p_2^2 - J_B\, p_1^2 + \frac{1}{2} J_C
\,p_2^2 + J_D\, p_1\cdot p_2 \,+\,\frac{1}{2} J_E\, p_1^2 + \frac{1}{2} \left( 1+\frac{3\varepsilon}{4} \right)K_0 \right.\nonumber\\[2.0mm]
  	&&{\hspace{0.5cm}}+\left.\left.\left( \xi-1 \right)\left( -\frac{1}{2} J_A\, p_2^2 - \frac{1}{2} J_B\, p_1^2 + \frac{1}{2} K_0\right)\right]\right\}\,.
\end{eqnarray}
\noindent where each of the functions $I_A,\cdots,J_E, K_0$ depend on $p_1,\,p_2$ and are given in Appendix Eqs~(\ref{eqn:Jscalardef}, \ref{Iscalardef}) as well as (\ref{eqn:J0def},~\ref{eqn:CLdef},~\ref{eqn:K0def}) below. 
Collecting the terms together with common Lorentz and Dirac structure at one loop we obtain: 
\begin{eqnarray}\label{eqn:part1}\fl
  	\mbf{P_1}^{\mu\nu}&=& \GS{p}_1\left(p_2^{\mu } p_1^{\nu }-p_1^{\mu } p_2^{\nu }\right) \left[\left(-I_E\, p_1^2-I_D\, p_2^2+J_B\right) (\xi-1) -2 (J_B-J_D-J_E) \right]\nonumber\\[2.0mm]
%%%%
&+&\GS{p}_2\left(p_2^{\mu } p_1^{\nu }-p_1^{\mu }
     p_2^{\nu }\right) \left[2 (J_0-2J_A-J_B+J_C+J_D)\right.\nonumber\\[2.0mm]
&&  \qquad \left.+\frac{1}{2} (\xi -1) \left(-2 I_D\, p_1^2+I_E\,
     p_1^2-I_C\, p_2^2+J_0-2 J_B+2 I_D\, p_1\cdot p_2\right) \right]\nonumber\\[2.0mm]
%%%%
&+&\GS{p}_1\GS{p}_2\left(\gamma ^{\nu } p_2^{\mu }+\gamma ^{\mu }p_1^{\nu }-\gamma ^{\mu } p_2^{\nu }-\gamma ^{\nu } p_1^{\mu }  \right) \left[J_0-J_A-J_B\right.\nonumber\\[2.0mm]
&&\qquad\left.+\frac{1}{4} (\xi -1) \left(I_E\, p_1^2+I_C\, p_2^2+J_0-2 J_A-2 J_B+2 I_D\, p_1\cdot
     p_2\right)\right]\nonumber\\[2.0mm]
%%%%
&+&\left(\gamma ^{\nu } p_2^{\mu }+\gamma ^{\mu } p_1^{\nu }-\gamma ^{\nu } p_1^{\mu }-\gamma ^{\mu } p_2^{\nu } \right) 
\left[\frac{1}{2} \left(2 J_B\, p_1^2-J_E\, p_1^2+2 J_A\, p_2^2-J_C\, p_2^2\right.\right.\nonumber\\[2.0mm]
&&{\hspace{2.8cm}}\left.-K_0-2 (2J_0-2 J_A-2 J_B+J_D)\, p_1\cdot p_2+3\right)\nonumber\\[4.0mm]
&&\qquad-\frac{1}{2} (\xi - 1) \left(-J_B\, p_1^2-J_A\, p_2^2+2 I_D\, (p_1\cdot p_2)^2\right.\nonumber\\[2.0mm]
&&{\hspace{2.8cm}}\left.\left.+K_0+\left(I_E\, p_1^2+I_C\, p_2^2+J_0-2 J_A-2J_B\right) \, p_1\cdot p_2\right)\right]\;.\nonumber\\
\end{eqnarray}
Except where explicitly stated otherwise, $K_0,\cdots,J_E$ are functions of $p_1$, $p_2$ as given in  the Appendix (\ref{eqn:Jscalardef},~\ref{Iscalardef}) as well as (\ref{eqn:J0def},~\ref{eqn:CLdef},~\ref{eqn:K0def}) below.
This answer, (\ref{eqn:part1}), can then be substituted into the first part of the transverse Ward-Takahashi relation, (\ref{eqn:twtr},~\ref{eqn:twtrcorrect}), {\it i.e.} in (\ref{eqn:twtrassum}).

%%\newpage
\subsection{Part 2}
\noindent It is useful here to note that the many integrals appearing in evaluation of the loop graphs are expressible in terms of a set of  basis functions $J_0$, $L$, $L'$, $\,\mathcal{C}$ and $S$ given by:
\vspace{1.5mm}
\begin{eqnarray}
 	J_0= \frac{2}{\Delta}\left[ Sp\left( \frac{p_1^2-p_1\cdot p_2+\Delta}{p_1^2} \right)-Sp\left( \frac{p_1^2-p_1\cdot p_2-\Delta}{p_1^2}\right)\right.\nonumber\\[2.mm]
  	 \left.\qquad\qquad\qquad\quad+\frac{1}{2}\ln\left( \frac{p_1\cdot p_2-\Delta}{p_1\cdot p_2+\Delta} \right)\ln\left( \frac{\left( p_2-p_1 \right)^2}{p_1^2} \right) \right]\label{eqn:J0def}\\[2.mm]
{\rm with}\quad\Delta=\,\sqrt{\left(p_1 \cdot p_2\right)^2\,-\,p_1^2 p_2^2}\quad ,\qquad {\rm and}\nonumber\\[2.mm]
  	L= \ln\left( -\frac{p_1^2}{\mu^2} \right)\quad;\qquad	L'= \ln\left( -\frac{p_2^2}{\mu^2} \right)\quad;\qquad S= \frac{1}{2}\ln\left( -\frac{(p_1-p_2)^2}{\mu^2} \right)\nonumber\\[2.mm]
 	\mathcal{C}= -\frac{2}{\varepsilon}-\gamma-\ln(\pi)\quad .\label{eqn:CLdef}
\end{eqnarray}

\noindent In terms of these the $K$-scalar integrals which are about to appear are given simply by: 
\begin{eqnarray}\label{eqn:K0def}
  	K_0(p_1,p_2)&=&2\left( \mathcal{C}+2-2S \right)\nonumber\\[1.5mm]
  	K_0(0,p_2)&=&2\left( \mathcal{C}+2-L' \right)\nonumber\\[1.5mm]
  	K_0(p_1,0)&=&2\left( \mathcal{C}+2-L \right)\quad .
\end{eqnarray}

\noindent With these definitions the $\left(S^{-1}_F(p_1)\sigma^{\mu\nu}+\sigma^{\mu\nu}S^{-1}_F(p_2)\right)$ piece of (\ref{eqn:twtrcorrect}) requires the evaluation of the inverse fermion propagator to one-loop. We note that the inverse propagator can be written as $S_F^{-1}(p_1)=\GS{p}_1-\Sigma_{(2)}(p_1)$, where
\begin{eqnarray}
  	\Sigma_{(2)}(p_1)&=& -\frac{i\alpha}{4\pi^3}\int d^d k \frac{\gamma^\alpha\left( \GS{p}_1-\GS{k} \right)\gamma^\beta}{k^2\left( p_1-k \right)^2}\left( g_{\alpha\beta}+\left( \xi-1 \right)\frac{k_\alpha k_\beta}{k^2} \right)\nonumber\\[3.mm]
  	&=&-\frac{i\alpha}{4\pi^3}\left( -2-\varepsilon \right)\left[ \GS{p}_1 \mbf{K^{(0)}}\left( p_1,0 \right)-\gamma^\lambda \mbf{K^{(1)}_\lambda}\left( p_1,0 \right) \right]\nonumber\\[3.mm]
  	&&-\,\frac{i\alpha}{4\pi^3}\left( \xi-1 \right)\left[\gamma^\lambda\GS{p}_1\gamma^\nu \mbf{J^{(2)}_{\lambda\nu}}\left( p_1,0 \right)+\gamma^\lambda \mbf{K^{(1)}_\lambda}(p_1,0)\right]\;.
\end{eqnarray}

\noindent Once the tensor integrals have been substituted, the inverse propagator is
%\vspace{2.5mm}
\begin{equation}\label{eqn:fermprop}
  	S_F^{-1}\left( p_1 \right)=\GS{p}_1\left( 1+\frac{\alpha\xi}{4\pi}\left( \mathcal{C}+1-L \right) \right)\; ,
\end{equation}

\noindent with a similar expression for $S_F^{-1}(p_2)$ with the replacement $p_1\rightarrow p_2$ in (\ref{eqn:fermprop}).
From these we deduce that
\begin{eqnarray}\label{eqn:part2}
  	\mbf{P_2}^{\mu\nu}&=& \GS{p}_1\left(\gamma ^{\mu }\gamma ^{\nu }-\gamma^\nu\gamma^\mu\right)\,\frac{\xi}{4}\,  \left(K_0(p_1,0)-2\right) \nonumber\\[2.mm]
%%%%
&+&\left(\GS{p}_2\left(\gamma ^{\mu }\gamma ^{\nu}-\gamma^\nu\gamma^\mu\right)
+\,4\gamma ^{\mu } p_2^{\nu }- 4\gamma ^{\nu } p_2^{\mu}\right)\,\frac{\xi}{4} \, \left(K_0(0,p_2)-2\right)\; . 
\end{eqnarray}
\noindent Substituting for these in the transverse Ward-Takahashi relation, (\ref{eqn:twtr},~\ref{eqn:twtrcorrect}), will give us the second piece.

\subsection{Part 3}
\noindent The third part we wish to compute involves the anticommutator of $\sigma^{\mu\nu}$ with the vector vertex with momentum routing as defined in figure~2. In the work of He~\cite{He:2002jg} the anticommutator is replaced by its $4$-dimensional identity, hence the appearance of the axial vector therein. We choose not to follow this path for obvious reasons --- the integrals must be evaluated in $d=4+\varepsilon$ dimensions.

\noindent Thus we begin with
\begin{eqnarray}
	-\frac{1}{2}\left( p_1+p_2 \right)_\lambda\,\left\{ \sigma^{\mu\nu},\Gamma_V^\lambda \right\}	&=& \frac{1}{2}\frac{i\alpha}{4\pi^3}\int d^4k\, \frac{\gamma^\alpha\left( \GS{p}_1-\GS{k} \right)\left\{\sigma^{\mu\nu},\gamma^\lambda\right\}\left( \GS{p}_2-\GS{k} \right)\gamma^\beta}{k^2\left( p_1-k \right)^2\left( p_2-k \right)^2}\nonumber\\[3.5mm]
  	&&\qquad\qquad\times\left( g_{\alpha\beta}+\left( \xi-1 \right)\frac{k_\alpha k_\beta}{k^2} \right)\;.
\end{eqnarray}

\noindent After extensive use of Dirac algebra identities and noting the definition of $r^{\lambda\mu\nu}$ in (\ref{eqn:chisholm}), we deduce 
%\vspace{3.5mm}
\begin{eqnarray}\fl
  	-\frac{1}{2}\left( p_1+p_2 \right)_\lambda\,\left\{ \sigma^{\mu\nu},\Gamma_V^\lambda \right\}\nonumber\\[3.5mm]
  	\fl= -\frac{i}{2}\left( p_1+p_2 \right)_\lambda\, r^{\lambda\mu\nu}\nonumber\\[3.5mm]
  	\fl+\,\frac{\alpha}{4\pi^3}\left( p_1+p_2 \right)_\lambda\left\{
  	 \left(\gamma^\beta r^{\lambda\mu\nu} \gamma^\delta\right)
  	\left(p_{1\delta}p_{2\beta}
  	\mbf{J^{(0)}}(p_1,p_2)-p_{1\delta}\mbf{J^{(1)}_\beta}(p_1,p_2)
	-\,p_{2\beta}\mbf{J^{(1)}_\delta}(p_1,p_2)\right)\right.\nonumber\\[3.5mm]
	{\hspace{10mm}}\,-\,\frac{1}{2}\gamma^\alpha\gamma^\beta\, r^{\lambda\mu\nu}\,\gamma^\delta\gamma_\alpha\mbf{J^{(2)}_{\beta\delta}}(p_1,p_2)\nonumber\\[3.5mm]
  	%\
  	{\hspace{10mm}}\,+\,\left( \xi-1 \right)\left[ \frac{1}{2}\gamma^\beta\GS{p}_1 r^{\lambda\mu\nu} \left(\mbf{J_\beta^{(1)}}(p_1,p_2)-\GS{p}_2\gamma^\delta\mbf{I_{\beta\delta}^{(2)}}(p_1,p_2)\right)\right. \nonumber\\[3.5mm]
{\hspace{29mm}}\left.\left.+\frac{1}{2} r^{\lambda\mu\nu}  \left( \GS{p}_2\gamma^\beta \mbf{J_\beta^{(1)}}(p_1,p_2)-\mbf{K^{(0)}}(p_1,p_2) \right)\right]\right\}\;.
\end{eqnarray}
\newpage
\noindent Substituting the forms of the integrals from the Appendix, we again collect the terms according to their Lorentz and Dirac structures
\begin{eqnarray}\label{eqn:part3}\fl
  	\mbf{P_3}^{\mu\nu}
= \GS{p}_1 \left[\left(I_E\, p_1^2-I_D\, p_2^2-J_B\right) (\xi
     -1) \left(p_2^{\mu } p_1^{\nu }-p_1^{\mu } p_2^{\nu }\right)-2
     (J_B+J_D-J_E) \left(p_2^{\mu } p_1^{\nu }-p_1^{\mu }
     p_2^{\nu }\right)\right]\nonumber\\[2.mm]
%%%%
\fl\hspace{10mm}\;+\GS{p}_1\GS{p}_2\left(\gamma ^{\nu } p_2^{\mu } - \gamma ^{\mu } p_2^{\nu } \right)\left[J_0+J_A-J_B-2 J_C+2J_D\right.\nonumber\\[2.mm]
%%%%
\fl\hspace{10mm}\;\left.\qquad+\frac{1}{4} (\xi -1) \left(4 I_D\, p_1^2+I_E\, p_1^2-3 I_C\, p_2^2+J_0+2 J_A-2 J_B+2 I_D\, p_1\cdot
     p_2\right)\right]\nonumber\\[2.mm]
%%%%
\fl\hspace{10mm}\;+   \GS{p}_2\left(p_2^{\mu } p_1^{\nu }-p_1^{\mu } p_2^{\nu }\right) \left[2 (J_0-J_B-J_C+J_D)\right.\nonumber\\[2.mm]
%%%%
\fl\hspace{10mm}\;\left.\qquad   +\frac{1}{2} (\xi
     -1) \left(2 I_D\, p_1^2+I_E\, p_1^2-I_C\, p_2^2+J_0-2
     J_B+2 I_D\, p_1\cdot p_2\right) \right]\nonumber\\[2.mm]
%%%%
\fl\hspace{10mm}\;+   \GS{p}_1\GS{p}_2\left(\gamma ^{\mu }p_1^{\nu }-\gamma ^{\nu } p_1^{\mu } \right) \left[J_0-J_A+J_B+2 J_D-2 J_E\right.\nonumber\\[2.mm]
%%%%
\fl\hspace{10mm}\;\left.\qquad+\frac{1}{4} (\xi -1) \left(-3 I_E\, p_1^2+I_C\, p_2^2+4 I_D\, p_2^2+J_0-2 J_A+2 J_B+2 I_D\, p_1\cdot p_2\right)\right]\nonumber\\[2.mm]
%%%%
\fl\hspace{10mm}\;+   \frac{1}{2}\GS{p}_2\left(\gamma ^{\mu }\gamma ^{\nu }-\gamma^\nu\gamma^\mu\right) \left[\frac{1}{2} \left(-2 J_0\, p_1^2+2
     J_A\, p_1^2-4 J_D\, p_1^2+3 J_E\, p_1^2+2 J_A\,
     p_2^2-J_C\, p_2^2\right.\right.\nonumber\\[1.mm]
%%%%
\fl\hspace{10mm}\;\left.\left.\,\,-K_0+2 (2 J_A-2 J_C+J_D) p_1\cdot
     p_2-5\right)+\frac{1}{4} (1-\xi ) \left(-3 I_E\, p_1^4+I_C\, p_2^2
     p_1^2+4 I_D\, p_2^2 p_1^2\right.\right.\nonumber\\[2.mm]
%%%%
\fl\hspace{10mm}\;\left.\left.\qquad\qquad+J_0\, p_1^2-2 J_A\, p_1^2-2
     J_A\, p_2^2+2 K_0-2 \left(I_D\, p_1^2-2 I_C\, p_2^2+2
     J_A\right)\, p_1\cdot p_2\right)\right]\nonumber\\[2.mm]
%%%%
\fl\hspace{10mm}\;+  \frac{1}{2}\GS{p}_1\left(\gamma ^{\mu }\gamma ^{\nu } -\gamma^\nu\gamma^\mu\right)\left[\frac{1}{2}
     \left(2 J_B\, p_1^2-J_E\, p_1^2-2 J_0\, p_2^2+2 J_B\,
     p_2^2+3 J_C\, p_2^2-4 J_D\, p_2^2-K_0\right.\right.\nonumber\\[2.mm]
%%%%
\fl\hspace{10mm}\;\left.\left.\qquad+2 (2 J_B+J_D-2
     J_E) p_1\cdot p_2-5\right)+\frac{1}{4} (1-\xi ) \left(-3 I_C\, p_2^4+4
     I_D\, p_1^2 p_2^2+I_E\, p_1^2 p_2^2\right.\right.\nonumber\\[1.mm]
%%%%
\fl\hspace{10mm}\;\left.\left.\qquad +J_0\, p_2^2
	-2J_B\, p_2^2-2 J_B\, p_1^2+2 K_0-2 \left(-2 I_E\,
     p_1^2+I_D\, p_2^2+2 J_B\right) p_1\cdot p_2\right)\right)\nonumber\\[2.mm]
%%%%
\fl\hspace{10mm}\;+   \left(\gamma ^{\mu } p_2^{\nu }-\gamma ^{\nu } p_2^{\mu } \right)
     \left[\frac{1}{2} \left(-2 J_B\, p_1^2-4 J_D\, p_1^2+3 J_E\, p_1^2+2
     J_A\, p_2^2-J_C\, p_2^2-K_0\right.\right.\nonumber\\[2.mm]
%%%%
\fl\hspace{10mm}\;\left.\left.\qquad+(4 J_0+4 J_A-4 J_B-8
     J_C+6 J_D) p_1\cdot p_2-5\right)\right.\nonumber\\[2.mm]
%%%%
\fl\hspace{10mm}\;\left.\qquad+\frac{1}{2} (1-\xi ) \left(-2 I_E\,
     p_1^4+2 I_D\, p_2^2 p_1^2+J_B\, p_1^2-J_A\, p_2^2-2
     I_D\, (p_1\cdot p_2)^2+K_0\right.\right.\nonumber\\[2.mm]
%%%%
\fl\hspace{10mm}\;\left.\left.\qquad\qquad-\left(4 I_D\, p_1^2+I_E\,
     p_1^2-3 I_C\, p_2^2+J_0+2 J_A-2 J_B\right) p_1\cdot
     p_2\right)\right]\nonumber\\[2.mm]
%%%% 
\fl\hspace{10mm}\;+   \left(\gamma^{\mu } p_1^{\nu }-\gamma^{\nu } p_1^{\mu }  \right)
	\left[\frac{1}{2} \left(2 J_B\, p_1^2-J_E\, p_1^2-4
     J_0\, p_2^2+2 J_A\, p_2^2+4 J_B\, p_2^2+3 J_C\, p_2^2-4
     J_D\, p_2^2\right.\right.\nonumber\\[2.mm]
%%%%
\fl\hspace{10mm}\;\left.\left.\qquad-K_0-2 (2 J_0-2 J_A-2 J_B+J_D) p_1\cdot
     p_2-5\right)+\frac{1}{2} (1-\xi ) \left(-I_C\, p_2^4+2 I_D\, p_1^2
     p_2^2\right.\right.\nonumber\\[2.mm]
%%%%
\fl\hspace{10mm}\;\left.\left.\qquad\qquad+I_E\, p_1^2 p_2^2+J_0\, p_2^2-J_A\, p_2^2-2
     J_B\, p_2^2-J_B\, p_1^2+2 I_D\, (p_1\cdot
     p_2)^2+K_0\right.\right.\nonumber\\[2.mm]
%%%%
\fl\hspace{10mm}\;\left.\left.\qquad+\left(I_E\, p_1^2+I_C\, p_2^2+2 I_D\,
     p_2^2+J_0-2 J_A-2 J_B\right) p_1\cdot p_2\right)\right]\qquad .
\end{eqnarray}
%%\newpage
\subsection{Part 4}
The Wilson-line contribution comprises integration over several non-local diagrams in order to retain gauge invariance. To one-loop order these contributions are explicitly:
\begin{eqnarray}
\fl	&{\hspace{-1cm}} &\int \frac{d^d k}{\left( 2\pi \right)^d}\,k_\lambda\,\left\{\sigma^{\mu\nu},\tilde\Gamma_V^\lambda\right\} \nonumber\\%[3.0mm]
\fl	&=&-\frac{i\alpha}{4\pi^3}\int d^dk\; k_\lambda\,\frac{\gamma^\alpha\left(\GS{p}_1-\GS{k}\right)\left\{\sigma^{\mu\nu},\gamma^\lambda\right\}\left( \GS{p}_2-\GS{k} \right)\gamma^\beta}{k^2\left( p_1-k \right)^2\left( p_2-k \right)^2}\left( g_{\alpha\beta}+\left( \xi-1 \right)\frac{k_\alpha k_\beta}{k^2} \right)\nonumber\\[3.0mm]
 \fl 	&-&\frac{i\alpha}{4\pi^3}\int d^dk\; \left[ \frac{\gamma^\beta\left( \GS{p}_1-\GS{k} \right)\left\{\sigma^{\mu\nu},\gamma^\alpha\right\}}{k^2\left( p_1-k \right)^2}+\frac{\left\{\sigma^{\mu\nu},\gamma^\alpha\right\}\left( \GS{p}_2-\GS{k} \right)\gamma^\beta}{k^2\left( p_2-k \right)^2} \right]\left( g_{\alpha\beta}+\left( \xi-1 \right)\frac{k_\alpha k_\beta}{k^2} \right)\, .\nonumber\\\fl 
\end{eqnarray}

\noindent Computing these two pieces separately, we find the first to be
\begin{eqnarray}
\fl  		&&-\frac{\alpha}{4\pi^3}\int d^dk\; k_\lambda\frac{\gamma^\alpha\left(\GS{p}_1-\GS{k}\right)\left\{\sigma^{\mu\nu},\tilde\Gamma_V^\lambda\right\}\left( \GS{p}_2-\GS{k} \right)\gamma^\beta}{k^2\left( p_1-k \right)^2\left( p_2-k \right)^2}\left( g_{\alpha\beta}+\left( \xi-1 \right)\frac{k_\alpha k_\beta}{k^2} \right)\nonumber\\[2.7mm]
\fl&=&-\frac{\alpha}{2\pi^3}\left\{ -\frac{1}{2} \mbf{J_\lambda^{(1)}}(p_1,p_2)\,\gamma^\alpha\GS{p}_1\, r^{\lambda\mu\nu}\, \GS{p}_2\gamma_\alpha \right.\nonumber\\[2.7mm]\,
\fl&&{\hspace{13mm}}+\,\frac{1}{2}\mbf{J_{\lambda\delta}^{(2)}}(p_1,p_2)\gamma^\alpha\left( \GS{p}_1 \, r^{\lambda\mu\nu}\, \gamma^\delta +\gamma^\delta \,r^{\lambda\mu\nu}\, \GS{p}_2\right)\gamma_\alpha\,-\,\left.\frac{1}{2}\mbf{K_\lambda^{(1)}}(p_1,p_2)\,\gamma^\alpha \,r^{\lambda\mu\nu}\, \gamma_\alpha\right\}\nonumber\\[2.7mm]
%
%  The ( \xi - 1 )-piece of L1
%
\fl  	&&+\,\frac{\alpha}{4\pi^3}\left( \xi-1 \right)\left\{
  	\mbf{K^{(1)}_\lambda}(p_1,p_2)\left( r^{\lambda\mu\nu} \right)
  	+\mbf{J^{(1)}_\lambda}(p_1,p_2)\left( -\GS{p}_1\gamma^\mu\gamma^\nu\GS{p}_2\gamma^\lambda+\gamma^\lambda\GS{p}_1\gamma^\nu\gamma^\mu\GS{p}_2 \right)\right.\nonumber\\[2.7mm]
\fl  	&&{\hspace{23mm}}+\,\mbf{J^{(2)}_{\lambda\alpha}}(p_1,p_2)\left( -\gamma^\alpha\GS{p}_1 r^{\lambda\mu\nu} - r^{\lambda\mu\nu}\GS{p}_2\gamma^\alpha
  	+\gamma^\alpha\gamma^\mu\gamma^\nu\GS{p}_2\gamma^\lambda-\gamma^\alpha\GS{p}_1\gamma^\nu\gamma^\mu\gamma^\lambda\right)\nonumber\\[2.7mm]
\fl  	&&{\hspace{23mm}}+\,\mbf{I^{(2)}_{\lambda\alpha}}(p_1,p_2)\left( p_1^2\gamma^\alpha\gamma^\mu\gamma^\nu\GS{p}_2\gamma^\lambda-\gamma^\alpha\GS{p}_1\gamma^\nu\gamma^\mu\gamma^\lambda p_2^2 \right)\nonumber\\[2.7mm]
\fl  	&&{\hspace{23mm}}+\,\left.\mbf{J_{\lambda\alpha}^{(2)}}(p_1,0)\left( \gamma^\lambda\GS{p}_1\gamma^\nu\gamma^\mu\gamma^\alpha \right)
+\mbf{J_{\lambda\alpha}^{(2)}}(0,p_2)\left( -\gamma^\alpha\gamma^\mu\gamma^\nu\GS{p}_2\gamma^\lambda \right)\right\}\;.
\end{eqnarray}

\noindent The second set of contributions to the Wilson-line at one-loop is:
\begin{eqnarray}
\fl  	&&-\frac{i\alpha}{4\pi^3}\int d^dk\; \left[ \frac{\gamma^\beta\left( \GS{p}_1-\GS{k} \right)\left\{\sigma^{\mu\nu},\gamma^\alpha\right\}}{k^2\left( p_1-k \right)^2}+\frac{\left\{\sigma^{\mu\nu},\gamma^\alpha\right\}\left( \GS{p}_2-\GS{k} \right)\gamma^\beta}{k^2\left( p_2-k \right)^2} \right]\left( g_{\alpha\beta}+\left( \xi-1 \right)\frac{k_\alpha k_\beta}{k^2} \right)\nonumber\\[2.7mm]
  	%%%%%%
\fl  	&=&-\frac{\alpha}{2\pi^3}\left\{ 
\left( p_{2\lambda}\mbf{K^{(0)}}(0,p_2)-p_{1\lambda}\mbf{K^{(0)}}(p_1,0)+\mbf{K^{(1)}_\lambda}(p_1,0)-\mbf{K^{(1)}_\lambda}(0,p_2)\right)
\right.\nonumber\\[2.7mm]
\fl&&{\hspace{45mm}}\times\left( 2\gamma^\mu g^{\lambda\nu}-2\gamma^\nu g^{\lambda\mu}+\varepsilon\left( \gamma^\mu\gamma^\nu-g^{\mu\nu} \right)\gamma^{\lambda} \right)\nonumber\\[2.7mm]
\fl&&{\hspace{45mm}}\left.	+\varepsilon\left( p_{1\lambda}\mbf{K^{(0)}}(p_1,0) -\mbf{K^{(1)}_\lambda}(p_1,0)\right)\, r^{\lambda\mu\nu}		\right\}\nonumber\\[2.7mm]
  	%
%
%	\xi-1 contribution
%
\fl  	&&+\,\frac{\alpha}{4\pi^3}\left( \xi-1 \right)\left\{
  	\mbf{J^{(2)}_{\lambda\alpha}}(p_1,0)\gamma^\alpha\GS{p}_1 \,r^{\lambda\mu\nu}  +\mbf{J^{(2)}_{\lambda\alpha}}(0,p_2)\,r^{\lambda\mu\nu}\GS{p}_2\gamma^\alpha\right.\nonumber\\[2.7mm]
\fl  	&&{\hspace{45mm}}-\left.\left(\mbf{K^{(1)}_\lambda}(p_1,0)+\mbf{K^{(1)}_\lambda}(0,p_2)\right)\,r^{\lambda\mu\nu} \right\}\; .
\end{eqnarray}
Collecting these together and again ordering the terms according to their Lorentz and Dirac structure, we have:
\begin{eqnarray}\label{eqn:part4}
\fl  	\mbf{P_4}^{\mu\nu}= \left(\GS{p}_1\GS{p}_2\gamma ^{\nu } p_1^{\mu }-\GS{p}_1\GS{p}_2\gamma^{\mu } p_1^{\nu }\right)\left[2 (J_B+J_D-J_E)+\left(-I_E\, p_1^2+I_D\, p_2^2+J_B\right) (\xi -1)\right] \nonumber\\[3.6mm]
%%%%
\fl\hspace{10mm}\; +\left(\GS{p}_1\GS{p}_2\gamma ^{\mu } p_2^{\nu }-\GS{p}_1\GS{p}_2\gamma ^{\nu } p_2^{\mu }\right)\left[2 (J_A-J_C+J_D)+\left(I_D\,
     p_1^2-I_C\, p_2^2+J_A\right) (\xi -1)\right] \nonumber\\[3.6mm]
%%%%
\fl\hspace{10mm}\;     +\GS{p}_2 \left[-2 I_D\, (\xi -1)
     \left(p_2^{\mu } p_1^{\nu }-p_1^{\mu } p_2^{\nu }\right) p_1^2-4
     (J_A-J_C) \left(p_2^{\mu } p_1^{\nu }-p_1^{\mu } p_2^{\nu
     }\right]\right)\nonumber\\[3.6mm]
%%%%
\fl\hspace{10mm}\;     +\GS{p}_1 \left[4 J_D \left(p_2^{\mu } p_1^{\nu
     }-p_1^{\mu } p_2^{\nu }\right)+2 \left(J_B-I_E\, p_1^2\right) (\xi -1)
     \left(p_2^{\mu } p_1^{\nu }-p_1^{\mu } p_2^{\nu }\right)\right]\nonumber\\[3.6mm]
%%%% 
\fl\hspace{10mm}\;+ \left(\gamma ^{\nu } p_2^{\mu }-\gamma ^{\mu } p_2^{\nu }    \right) \left[-2 J_D\, p_1^2+2
     J_C\, p_2^2-K_0+4 (J_A-J_C+J_D\,)\, p_1\cdot p_2\right.\nonumber\\[3.6mm]
\fl\hspace{10mm}\;     \qquad+\frac{1}{2} (1-\xi ) \left(-2 I_E\, p_1^4+2 I_D\, p_2^2
     p_1^2+J_E\, p_1^2-3 J_C\, p_2^2+2 K_0\right.\nonumber\\[3.6mm]
\fl\hspace{10mm}\;     \left.\left.\qquad\qquad-2 \left(2 I_D\,
     p_1^2-2 I_C\, p_2^2+2 J_A+J_D\right) p_1\cdot
     p_2-3\right)+2\right]\nonumber\\[3.6mm]
%%%% 
\fl\hspace{10mm}\;-\frac{1}{2}\GS{p}_1 \left(\gamma ^{\mu }\gamma ^{\nu } -\gamma^\nu\gamma^\mu\right)\left[J_E\, p_1^2-2 J_A\, p_2^2+J_C\,
     p_2^2+2 J_D\, p_1\cdot p_2 - 4\right.\nonumber\\[3.0mm]
\fl\hspace{10mm}\;     \qquad\left.+\frac{1}{2} (1-\xi )\left(-2 I_C\,
     p_2^4+2 I_E\, p_1^2 p_2^2+2 J_A\, p_2^2-4 J_D\,
     p_2^2+K_0-4 J_E\, p_1\cdot p_2-4\right)-1\right]\nonumber\\[3.6mm]
%%%%
\fl\hspace{10mm}\;+ \frac{1}{2}\GS{p}_2\left(\gamma ^{\mu }\gamma ^{\nu }-\gamma^\nu\gamma^\mu\right) \left[2
     J_B\, p_1^2-J_E\, p_1^2-J_C\, p_2^2-2 J_D\, p_1\cdot
     p_2\right.\nonumber\\[3.6mm]
\fl\hspace{10mm}\;\qquad+\frac{1}{4} (\xi -1) \left(-3 I_E\, p_1^4+I_C\, p_2^2 p_1^2+4
     I_D\, p_2^2 p_1^2+J_0\, p_1^2-4 J_D\, p_1^2+J_E\,
     p_1^2-3 J_C\, p_2^2\right.\nonumber\\[3.mm]
\fl\hspace{10mm}\;\left.\left.\qquad\qquad+3 K_0-K_0(0,p_2)-2 \left(I_D\, p_1^2-2
     I_C\, p_2^2+2 J_A+2 J_C+J_D\right) p_1\cdot
     p_2-5\right)+1\right]\nonumber\\[3.6mm]
%%%% 
\fl\hspace{10mm}\;+ \left(\gamma ^{\mu } p_1^{\nu } -\gamma^\nu p_1^\mu\right)\left[4 J_A\, p_2^2-2J_C\, p_2^2-2 J_D\, p_2^2-K_0+4 (J_B-J_E) p_1\cdot p_2+4\right.\nonumber\\[3.6mm]
\fl\hspace{10mm}\;\qquad+\frac{1}{2} (\xi -1) \left(-I_C\, p_2^4+2 I_D\, p_1^2 p_2^2+I_E\, p_1^2 p_2^2+J_0\, p_2^2-4 J_D\, p_2^2+K_0\right.\nonumber\\[3.6mm]
\fl\hspace{10mm}\;\left.\left.\qquad\qquad-K_0(p_1,0)+\left(2 I_D\, p_2^2-4 J_E\right) p_1\cdot p_2-2\right)+4\right]\nonumber\\[3.6mm]
\fl\hspace{10mm}\;+ \frac{1}{2}\GS{p}_1\left(\gamma ^{\mu }\gamma^{\nu }-\gamma^\nu\gamma^\mu\right)	\left[\frac{1}{2} \left(4-K_0(p_1,0)\right) (\xi -1)-2\right] \nonumber\\[3.6mm]
%%%%
\fl\hspace{10mm}\;+\frac{1}{2}\GS{p}_2\left(\gamma ^{\mu}\gamma ^{\nu }-\gamma^\nu\gamma^\mu\right)\left[\frac{1}{2} \left(4-K_0(0,p_2)\right) (\xi -1)+2\right] \nonumber\\[3.6mm]
%%%%
\fl\hspace{10mm}\;    +\left(\gamma ^{\mu } p_1^{\nu} -\gamma ^{\nu } p_1^{\mu }  \right)\left[K_0(p_1,0)+\xi -5\right] \nonumber\\[3.6mm]
%%%%
\fl\hspace{10mm}\;     +\left( \gamma ^{\nu } p_2^{\mu } -  \gamma ^{\mu } p_2^{\nu }  \right)\left[1 + \xi K_0(0,p_2)\right] \quad .
\end{eqnarray}

\section{Summing terms gives the result}

We now collect together the terms from (\ref{eqn:part1},~\ref{eqn:part2},~\ref{eqn:part3},~\ref{eqn:part4}) to form
\begin{equation}\label{eqn:checkPsum}
\mbf{Q}^{\mu\nu}\;=\;\left[ \mbf{P_1} - \mbf{P_2} - \mbf{P_3} - \mbf{P_4} \right]^{\mu\nu}\;.
\end{equation}
The fact that each term is individually antisymmetric in $\mu$ and $\nu$ assists the checking. Rather than report every term of this somewhat tedious exercise,
we illustrate that the answer for $\mbf{Q}^{\mu\nu}$ is indeed zero by considering two structures. First those proportional to $\mathcal{C}$ of (\ref{eqn:CLdef}), which are individually singular in $1/\varepsilon$, then we will consider those proportional to unity.

%\newpage
\subsection{Equating $\mathcal{C}$'s}
%P1 $\xi=1$
We denote the singular term in $\mbf{Q}^{\mu\nu}$ by $\mbf{Q_{\varepsilon}}^{\mu\nu}$. Collecting these  parts we have:
\begin{eqnarray}
\mbf{Q_{\varepsilon}}^{\mu\nu}&\equiv&\gamma ^{\mu }
     \left(p_2^{\nu }-p_1^{\nu }\right) \,\mathcal{C}
%P1 $\xi-1$
\;+\;	(\xi -1)\,\gamma
     ^{\mu } \left(p_2^{\nu }-p_1^{\nu }\right)\,\mathcal{C} 
{\hspace{29.1mm}}(+\mbf{P_1}^{\mu\nu})
\nonumber\\[2.mm]
%P2 $\xi=1$
&&
%-	\frac{1}{2}\left(\GS{p}_1+\GS{p}_2\right)\,\gamma ^{\mu }\gamma ^{\nu }\,\mathcal{C}\,-\,2 \gamma ^{\mu } p_2^{\nu } \,\mathcal{C}
%P2 $\xi-1$
%&&
-  \xi\,\left[ \frac{1}{2}\,\left(\GS{p}_1+\GS{p}_2\right)\,\gamma ^{\mu }\gamma ^{\nu }\,
  	+\,2 \gamma ^{\mu } p_2^{\nu }\right] \,\mathcal{C}
{\hspace{36.7mm}}(-\mbf{P_2}^{\mu\nu})\nonumber\\[2.mm]
%P3 $\xi=1$
&&
%	+ \frac{1}{2}\,\left(\GS{p}_1+\GS{p}_2\right)\,\gamma ^{\mu }\gamma ^{\nu }\,\mathcal{C}\,-\,\gamma ^{\mu }
%     \left(p_1^{\nu }+p_2^{\nu }\right) \,\mathcal{C}
%P3 $\xi-1$
%&&
	+ \xi\,\left[ \frac{1}{2}\,\left(\GS{p}_1+\GS{p}_2\right)\,\gamma ^{\mu }\gamma ^{\nu }
  	\,+\,\gamma ^{\mu }
     \left(p_1^{\nu }+p_2^{\nu }\right)\right] \,\mathcal{C}
{\hspace{26.7mm}}(-\mbf{P_3}^{\mu\nu})
\nonumber\\[2.mm]
%WL1 $\xi=1$
&&+	2\,\gamma ^{\mu }
     \left(p_2^{\nu }-p_1^{\nu }\right)\,\mathcal{C}\,+\,	2 \gamma ^{\mu }
     \left(p_1^{\nu }-p_2^{\nu }\right) \,\mathcal{C}\nonumber\\[2.mm]
%WL1 $\xi-1$
%&&
&&	+ (\xi -1) \,\left[ \frac{1}{2}\,\left(\GS{p}_1+\GS{p}_2\right)\,\gamma ^{\mu }\gamma ^{\nu }\,
  	+\,2 \gamma ^{\mu } p_2^{\nu }\right] \,\mathcal{C}\nonumber\\[2.mm]
%WL2 $\xi-1$
&&	-  (\xi -1)\, \left[\frac{1}{2}\,\left(\GS{p}_1+\GS{p}_2\right)\,\gamma ^{\mu }\gamma ^{\nu }
  	\,+2\, \gamma ^{\mu } p_2^{\nu }\right]\,\mathcal{C}
{\hspace{27.6mm}}(-\mbf{P_4}^{\mu\nu})\nonumber\\[2.mm]
&&\qquad\qquad-\qquad\left( \mu\leftrightarrow\nu \right)\nonumber\\[3.0mm] 
&=& \mbf{0}\qquad .
\end{eqnarray}
The $\mbf{P_i}^{\mu\nu}$ in brackets indicate from which part the terms originate.
\newpage

\subsection{Equating $1$'s}
We next collect together the terms proportional to unity. We denote this collection by $\mbf{Q_{(1)}}^{\mu\nu}$:
\begin{eqnarray}
  	\mbf{Q_{(1)}}^{\mu\nu}&=&\, \frac{1}{\Delta^2}\Bigg\{ (2-\xi)\left(-p_1^{\mu} p_2^{\nu}\right)
  	\left[\GS{p}_2 \left(p_1^{2}-p_1\cdot p_2\right)+\GS{p}_1 \left(p_2^{2}-p_1\cdot p_2\right)\right] \nonumber\\[2.mm]
&&+ 2\, (\xi -1)\, \gamma ^{\mu } 	\left(p_2^{\nu}-p_1^{\nu }\right)\,\Delta^2 {\hspace{50mm}}(+\mbf{P_1}^{\mu\nu})\nonumber\\[2.mm]
&&-\,\xi\,\left[\frac{1}{2}\left(\GS{p}_1+\GS{p}_2\right)\gamma^{\mu}\gamma^{\nu}
+2 \gamma ^{\mu } p_2^{\nu }\right]\, \Delta^2 {\hspace{35mm}}(-\mbf{P_2}^{\mu\nu})\nonumber\\[2.mm]
&&+\,\frac{(4 + \xi)}{2}\left(\GS{p}_1+\GS{p}_2\right)\gamma ^{\mu }\gamma ^{\nu }\,\Delta^2+6\gamma^\mu\left( p_1^\nu+p_2^\nu \right)\Delta^2\nonumber\\[2.mm]
&&-\,(2 - \xi)\,\left\{ \gamma^\mu p_1^\nu\left[ p_2^2\left( p_1\cdot p_2+p_1^2 \right)+2\Delta^2 \right]-\gamma^\mu p_2^\nu p_1^2\left( p_2^2+p_1\cdot p_2\right)\right.\nonumber\\[2.7mm]
&&\hspace{15mm}+\GS{p}_1\GS{p}_2\gamma^\mu\left[ p_2^\nu\left( p_1^2+p_1\cdot p_2 \right)-p_1^\nu\left( p_2+p_1\cdot p_2 \right) \right]\nonumber\\[2.mm]
&&\hspace{15mm}+\left.p_1^\mu p_2^\nu\left[ \GS{p}_2\left( p_1^2+p_1\cdot p_2 \right)-\GS{p}_1\left( p_2^2+p_1\cdot p_2 \right) \right]\right\}{\hspace{3.0mm}}(-\mbf{P_3}^{\mu\nu})\nonumber\\[2.mm]
&&-\left(2\GS{p}_1\gamma^\mu\gamma^\nu+4\gamma^\mu p_1^\nu\right)\Delta^2 -\left(2\GS{p}_2\gamma^\mu\gamma^\nu+4\gamma^\mu p_2^\nu\right)\Delta^2\nonumber\\[2.mm]
&&+\, (2-\xi)\,\left\{ \gamma^\mu p_1^\nu p_2^2\left( p_1^2+p_1\cdot p_2 \right)-\gamma^\mu p_2^\nu p_1^2\left( p_2^2+p_1\cdot p_2 \right)\right.\nonumber\\[2.mm]
&&\hspace{15mm}+\GS{p}_1\GS{p}_2\gamma^\mu\left[ p_2^\nu\left( p_1^2+p_1\cdot p_2 \right)-p_1^\nu\left( p_2^2+p_1\cdot p_2 \right) \right]\nonumber\\[2.mm]
&&\hspace{15mm}+\left.p_1^\mu p_2^\nu\left( 2\GS{p}_2p_1^2-2\GS{p}_1p_1\cdot p_2 \right)\right\}{\hspace{33.0mm}}(-\mbf{P_4}^{\mu\nu})\nonumber\\[2.mm]
&&\qquad\qquad-\qquad\left( \mu\leftrightarrow\nu \right)\qquad\Bigg\}\nonumber\\[2.0mm]
&=& \mbf{0}\qquad .
\end{eqnarray}

%\newpage
\subsection{The Result}

We can similarly check for  all the terms in $\mbf{Q}^{\mu\nu}$, {\it i.e.} for the combination of $\mbf{P_i}^{\mu\nu}$ $(i=1,4)$ of (\ref{eqn:checkPsum}), that the result is indeed zero.

\noindent Thus in conclusion we have deduced that the transverse Ward-Takahashi identity
to one loop order in $d$ close to 4 dimensions is indeed given by (\ref{eqn:twtrcorrect}):
\newpage
\begin{eqnarray}
\fl	\nonumber
  	iq^\mu\Gamma^\nu_V(p_1,p_2)-iq^\nu\Gamma^\mu_V(p_1,p_2)=S^{-1}_F(p_1)\,\sigma^{\mu\nu}+\sigma^{\mu\nu}\,S^{-1}_F(p_2)-\,\frac{1}{2}(p_1+p_2)_\lambda\left\{\sigma^{\mu\nu}, \Gamma^\lambda_V\right\}\nonumber\\[2.7mm]
	\hspace{27mm}+\int\frac{d^dk}{\left( 2\pi \right)^d}\;k_\lambda\left\{\sigma^{\mu\nu},\tilde\Gamma_V^\lambda\right\}\;.
\end{eqnarray}

\noindent 
This is a critical step in developing non-perturbative Feynman rules for the key fermion-boson interaction so central to Abelian and non-Abelian gauge theories in the strong coupling regime.
\vspace{0.9cm}

%\noindent {\bf Acknowledgements}
\ack
\noindent RW is grateful to the UK Particle Physics and Astronomy Research Council (PPARC) for the award of a research studentship.
We acknowledge partial support of the EU-RTN Programme, 
Contract No. HPRN-CT-2002-00311, \lq\lq EURIDICE''. We are most grateful to Gino Isidori for the opportunity to participate in the 2005 LNF Spring Institute at Frascati, where a key part of this work was done.

\vspace{0.6cm}

%%%%%%%%%%%%%%%%%%%%%%%%%%%%%%%
%					%
%    APPENDICES START HERE	%
%					%
%%%%%%%%%%%%%%%%%%%%%%%%%%%%%%%

%\newpage

%\baselineskip=6.mm
\setcounter{section}{0}
\renewcommand{\thesection}{Appendix \Alph{section}}
\section{}
\renewcommand{\thesection}{\Alph{section}}
\setcounter{equation}{0}
\renewcommand{\theequation}{\thesection.\arabic{equation}}

Note that in the text these functions will be assumed to depend upon $p_1$ and $p_2$ unless explicitly indicated.
\begin{eqnarray}
\fl  	\mbf{K^{(0)}}\left( p_1,p_2 \right)&=& \int d^4\omega\; \frac{1}{\left( p_1-\omega \right)^2\left( p_2-\omega \right)^2}{\hspace{10mm}}=\; \frac{i \pi^2}{2} K_0\left(p_1,p_2\right)\;,\\[1.5mm]
\fl  	\mbf{K^{(0)}}\left( 0,p_2 \right)&=& \int d^4\omega\; \frac{1}{\omega^2\left( p_2-\omega \right)^2}{\hspace{22mm}}=\; \frac{i \pi^2}{2} K_0\left(0,p_2\right)\;,\\[1.5mm]
\fl  	\mbf{K^{(0)}}\left( p_1,0 \right)&=& \int d^4\omega\; \frac{1}{\left( p_1-\omega \right)^2\omega^2}{\hspace{22mm}}=\; \frac{i \pi^2}{2} K_0\left(p_1,0\right)\;,\\[1.5mm]
\fl  	\mbf{K_\mu^{(1)}}\left( p_1,p_2 \right)&=& \int d^4\omega\; \frac{\omega_\mu}{\left( p_1-\omega \right)^2\left( p_2-\omega \right)^2}{\hspace{10mm}}=\;\frac{i \pi^2}{4}\left( p_1+p_2 \right)_\mu K_0\left(p_1,p_2\right)\;,\\[1.5mm]
\fl  	\mbf{K_\mu^{(1)}}\left( 0,p_2 \right)&=& \int d^4\omega\; \frac{\omega_\mu}{\omega^2\left( p_2-\omega \right)^2}{\hspace{22mm}}=\;\frac{i \pi^2}{4}p_{2\mu} K_0\left(0,p_2\right)\;,\\[1.5mm]
\fl  	\mbf{K_\mu^{(1)}}\left( p_1,0 \right)&=& \int d^4\omega\; \frac{\omega_\mu}{\left( p_1-\omega \right)^2\omega^2}{\hspace{22mm}}=\;\frac{i \pi^2}{4}p_{1\mu} K_0\left(p_1,0\right)\;,\\[1.5mm]
\fl  	\mbf{J^{(0)}}\left( p_1,p_2 \right)&=&\int d^4\omega\; \frac{1}{\omega^2\left( p_1-\omega \right)^2\left( p_2-\omega \right)^2}{\hspace{5mm}}=\;\frac{i \pi^2}{2} \,J_0\;,\\[1.5mm]
\fl  	\mbf{J_\mu^{(1)}}\left( p_1,p_2 \right)&=&\int d^4\omega\; \frac{\omega_\mu}{\omega^2\left( p_1-\omega \right)^2\left( p_2-\omega \right)}{\hspace{7mm}}=\;\frac{i \pi^2}{2}\left[ p_{2\mu}J_A+p_{1\mu}J_B \right]\;,\\[1.5mm]
\fl  	\mbf{J_\mu^{(1)}}\left( 0,p_2 \right)&=&\int d^4\omega\; \frac{\omega_\mu}{\omega^4\left( p_2-\omega \right)^2}
{\hspace{22mm}}=\;\frac{i \pi^2 p_{2\mu}}{p_2^2}\;,\\[1.5mm]
\fl  	\mbf{J_\mu^{(1)}}\left( p_1,0 \right)&=&\int d^4\omega\; \frac{\omega_\mu}{\omega^4\left( p_1-\omega \right)^2}
{\hspace{22mm}}=\;\frac{i \pi^2 p_{1\mu}}{p_1^2}\;,\\[1.5mm]
\fl  	\mbf{J_{\mu\nu}^{(2)}}\left( p_1,p_2 \right)&=&\int d^4\omega\; \frac{\omega_\mu\omega_\nu}{\omega^2\left( p_1-\omega \right)^2\left( p_2-\omega \right)^2}\nonumber\\[1.5mm]
\fl  		&=&\frac{i\pi^2}{2}\left\{ \frac{g_{\mu\nu}}{d}K_0 +\left( p_{2\mu}p_{2\nu}-g_{\mu\nu}\frac{p_2^2}{4} \right)J_C
  	+\left( p_{1\mu}p_{1\nu}-g_{\mu\nu}\frac{p_1^2}{4} \right)J_E\right.\nonumber\\[1.5mm]
\fl  	&&\left. +\left( p_{1\mu}p_{2\nu}+p_{2\mu}p_{1\nu}-g_{\mu\nu}\frac{p_1\cdot p_2}{2} \right)J_D  \right\}\;,\\[3mm]
\fl  	\mbf{J_{\mu\nu}^{(2)}}\left( 0,p_2 \right)&=&\int d^4\omega\; \frac{\omega_\mu\omega_\nu}{\omega^4\left( p_2-\omega \right)^2}\nonumber\\[1.5mm]
\fl  		&=&\frac{i\pi^2}{2}\left( \frac{g_{\mu\nu}}{4}K_0\left( 0,p_2 \right)+\frac{p_{2\mu}p_{2\nu}}{p_2^2} \right)\;,\\[1.5mm]
\fl  	\mbf{J_{\mu\nu}^{(2)}}\left( p_1,0 \right)&=&\int d^4\omega\; \frac{\omega_\mu\omega_\nu}{\omega^4\left( p_1-\omega \right)^2}\nonumber\\[1.5mm]
\fl  		&=&\frac{i\pi^2}{2}\left( \frac{g_{\mu\nu}}{4}K_0\left( p_1,0 \right)+\frac{p_{1\mu}p_{1\nu}}{p_1^2} \right)\;,\\[3mm]
\fl  	\mbf{I_\mu^{(1)}}\left( p_1,p_2 \right)&=&\int d^4\omega\; \frac{\omega_\mu}{\omega^4\left( p_1-\omega \right)^2\left( p_2-\omega \right)^2}\nonumber\\[1.5mm]
\fl  		&=&\frac{i \pi^2}{2}\left[ p_{2\mu}\,I_A+p_{1\mu}\,I_B\, \right]\;,\\[3mm]
\fl  	\mbf{I_{\mu\nu}^{(2)}}\left( p_1,p_2 \right)&=&\int d^4\omega\; \frac{\omega_\mu\omega_\nu}{\omega^4\left( p_1-\omega \right)^2\left( p_2-\omega \right)^2}\nonumber\\[1.5mm]
\fl  		&=&\frac{i\pi^2}{2}\left\{ \frac{g_{\mu\nu}}{4}\,J_0 +\left( p_{2\mu}p_{2\nu}-g_{\mu\nu}\frac{p_2^2}{4} \right)I_C
  	+\left( p_{1\mu}p_{1\nu}-g_{\mu\nu}\frac{p_1^2}{4} \right)I_E\right.\nonumber\\[1.5mm]
\fl  	&&\left. +\left( p_{1\mu}p_{2\nu}+p_{2\mu}p_{1\nu}-g_{\mu\nu}\frac{p_1\cdot p_2}{2} \right)I_D  \right\}\;.
\end{eqnarray}

\subsection{J-scalar functions}
\begin{eqnarray}\label{eqn:Jscalardef}
\fl  	J_A&=& \frac{1}{\Delta^2}\left\{ \frac{J_0}{2}\left( -p_1^2\left( p_2^2-p_1\cdot p_2 \right) \right)+p_1\cdot p_2\, L'-p_1^2L-2\left( p_1\cdot p_2-p_1^2 \right)S \right\}\nonumber\\[1.5mm]
\fl  	J_B&=& \frac{1}{\Delta^2}\left\{ \frac{J_0}{2}\left( -p_2^2\left( p_1^2-p_1\cdot p_2 \right) \right)+p_1\cdot p_2\, L-p_2^2L'-2\left( p_1\cdot p_2-p_2^2 \right)S \right\}\nonumber\\[1.5mm]
\fl  	J_C&=& \frac{1}{4\Delta^2}\left\{ 2p_1^2-4p_1\cdot p_2\, S+2p_1\cdot p_2\, L'+\left( 2p_1\cdot p_2\,p_1^2-3p_1^2p_2^2 \right)J_A-p_1^4J_B \right\}\nonumber\\[1.5mm]
\fl  	J_D&=& \frac{1}{4\Delta^2}\left\{ 2p_1\cdot p_2\,p_2^2J_A+2p_1\cdot p_2\,p_1^2J_B-2p_1\cdot p_2+2p_2^2S-p_2^2L'-p_1^2p_2^2J_A \right.\nonumber\\[1.5mm]
\fl  	&&{\hspace{32mm}}\left.+2p_1^2S-p_1^2L-p_1^2p_2^2J_B \right\}\nonumber\\[1.5mm]
\fl  	J_E&=& \frac{1}{4\Delta^2}\left\{ 2p_2^2-4p_1\cdot p_2\, S+2p_1\cdot p_2\, L+\left( 2p_1\cdot p_2\,p_2^2-3p_1^2p_2^2 \right)J_B-p_2^4J_A \right\}
\end{eqnarray}
\subsection{I-scalar functions}
\begin{eqnarray}\label{Iscalardef}
\fl  	I_A=\frac{1}{\Delta^2}\left\{ -\frac{\left(p_1^2-p_1\cdot p_2\right)}{2}\,J_0+2\left( 1-\frac{p_1\cdot p_2}{p_2^2} \right)S-L'+\frac{p_1\cdot p_2}{p_2^2}L \right\}\nonumber\\[1.5mm]
%%%%
\fl  	I_B=\frac{1}{\Delta^2}\left\{ -\frac{\left(p_2^2-p_1\cdot p_2\right)}{2}\,J_0+2\left( 1-\frac{p_1\cdot p_2}{p_1^2} \right)S-L+\frac{p_1\cdot p_2}{p_1^2}L'\right\}\nonumber\\[1.5mm]
%%%%
\fl  	I_C= \frac{1}{4\Delta^2}\left\{ 2p_1^2 J_0-4\frac{p_1\cdot p_2}{p_2^2}+\left( 2p_1\cdot p_2\,-3p_1^2 \right)J_A-p_1^2J_B-p_1^4\,I_B\right.\nonumber\\[1.5mm]
\fl  {\hspace{31mm}}\left.+\,\left( 2p_1\cdot p_2\, p_1^2-3p_1^2p_2^2 \right)I_A\, \right\}\nonumber\\[1.5mm]
%%%%
\fl  	I_D= \frac{1}{4\Delta^2}\left\{ -2p_1\cdot p_2\,J_0+4+\left( 2p_1\cdot p_2\,-p_2^2 \right)J_A+\left( 2p_1\cdot p_2\,-p_1^2 \right)J_B\right.\nonumber\\[1.5mm]
\fl  	{\hspace{31mm}}\left.+\,\left( 2p_1\cdot p_2\,p_2^2-p_1^2p_2^2 \right)I_A\,+\,\left( 2p_1\cdot p_2\, p_1^2 -p_1^2p_2^2\right)I_B \right\}\nonumber\\[1.5mm]
%%%%
\fl  	I_E= \frac{1}{4\Delta^2}\left\{ 2p_2^2 J_0-4\frac{p_1\cdot p_2}{p_1^2}+\left( 2p_1\cdot p_2\,-3p_2^2 \right)J_B-p_2^2J_A-p_2^4\,I_A\right.\nonumber\\[1.5mm]
\fl   {\hspace{31mm}}\left.+\,\left( 2p_1\cdot p_2\, p_2^2-3p_1^2p_2^2 \right)I_B\, \right\}\nonumber\\[1.5mm]
\end{eqnarray}

\end{document}